\documentclass[12pt]{iopart}
\usepackage[T1]{fontenc}
\usepackage[latin1]{inputenc}
\usepackage[english]{babel}
\usepackage{amssymb}
\usepackage{amsfonts}
\expandafter\let\csname equation*\endcsname\relax
\expandafter\let\csname endequation*\endcsname\relax
\usepackage{amsmath}
\usepackage{iopams}
\usepackage{amsopn}
\usepackage{graphicx}
\usepackage{color}
\usepackage{ulem}

\begin{document}

\title{Spin-correlations and magnetic structure in an Fe monolayer on 5$d$ transition metal surfaces}
\author{E. Simon$^1$, K. Palot\'as$^1$, B. Ujfalussy$^2$, A. De\'ak$^1$, G. M. Stocks$^3$ and L. Szunyogh$^{1,4}$}
\address{$^1$Department of Theoretical Physics, Budapest University of Technology and Economics, Budafoki \'{u}t 8., H-1111 Budapest, Hungary\\
$^2$Institute for Solid State Physics and Optics, Wigner Research Centre for Physics, Hungarian Academy of Sciences, P. O. Box 49, H-1525 Budapest, Hungary\\
$^3$Materials Science and Technology Division, Oak Ridge National Laboratory, Oak Ridge, TN 37831-6057, USA \\
$^4$Condensed Matter Research Group of Hungarian Academy of Sciences, Budapest University of Technology and Economics, Budafoki \'ut 8, H-1111 Budapest, Hungary}
\ead{esimon@phy.bme.hu}

\begin{abstract}
We present a detailed first principles study on the magnetic structure of an Fe monolayer on
different surfaces of 5$d$ transition metals. We use the spin-cluster expansion technique to obtain parameters of a spin model, and predict the possible magnetic ground state of the studied systems by employing the mean field approach and in certain cases by spin dynamics calculations. We point out that the number of shells
considered for the isotropic exchange interactions plays a crucial role in the determination of the magnetic ground state.
In the case of Ta substrate we demonstrate that the out-of-plane relaxation of the Fe monolayer causes a transition from ferromagnetic to antiferromagnetic ground state.
We examine the relative magnitude of nearest neighbour Dzyaloshinskii-Moriya ($D$) and isotropic ($J$) exchange interactions in order to get insight into the nature of magnetic pattern formations. For the Fe/Os(0001) system we calculate a very large $D/J$ ratio, correspondingly, a spin spiral ground state. We find that, mainly through the leading isotropic exchange and Dzyaloshinskii-Moriya interactions, the inward layer relaxation substantially influences the magnetic ordering of the Fe monolayer. For the Fe/Re(0001) system characterized by large antiferromagnetic interactions we also determine the chirality of the $120^{\circ}$ N\'eel-type ground state.
\end{abstract}


\maketitle

\section{Introduction}

Magnetic nano-clusters and magnetic thin films on non-magnetic substrates are in the focus of experimental and theoretical research since such systems play an ultimate role in technologically relevant magnetic devices.
The dependence of the magnetic ground state of ultrathin films on the quality of the underlying substrate is an interesting and relevant issue.
The study of temperature induced magnetic reversal in thin films and nano-structures is of paramount importance for hard disk and magnetic sensor technologies.
Ab initio electronic structure methods already proved to give a proper description of magnetic properties of solids, and these methods are often used to obtain parameters for spin models. Using spin model parameters the description of complex magnetic structures or finite temperature magnetism becomes feasible.

Fe thin films on 5$d$ transition metal surfaces have been extensively studied both experimentally and theoretically.
Whilst a $c(2\times2)$ antiferromagnetic (AFM) state was discovered for an
Fe monolayer (ML) on BCC W(001) surface \cite{kubetzka, qian}, on the BCC Ta$_{x}$W$_{1-x}$(001) surface a crossover was found between AFM (pure W substrate) and ferromagnetic (FM) (pure Ta substrate) states with the same Fe layer relaxation \cite{ferriani,ondracek}. On the other hand, previous theoretical studies demonstrated that the magnetic ground state strongly depends on the relaxation of the magnetic overlayer. In case of an Fe ML on FCC Ir(001), the large layer relaxation leads to a spin spiral ground state \cite{adeak,kudrnovsky}. An investigation of the magnetism of an Fe ML on Pt(111) and at a Pt step-edge also led to the conclusion that the crystallographic structure plays a significant role in the formation of the magnetic structure \cite{repetto}. Another mechanism that can affect the magnetic ordering of a monolayer is the hybridization between the monolayer and the substrate. Hardrat et al.\ demonstrated the change of the exchange coupling and the magnetic ordering of an Fe ML on HCP(0001) and FCC(111) surfaces of 4$d$ and 5$d$ transition metals due to the change in $d$-band filling \cite{hardrat}.
When increasing the atomic number of the substrate element, the $d$-band of the substrate moves downwards with respect to the Fermi level while the energetic position of the Fe 3$d$ band remains roughly unchanged. Due to corresponding changes in the hybridization between the substrate 4$d$ or 5$d$ band and the Fe 3$d$ band, the Fe spin moment gradually increases towards the brinks of the 4$d$ or 5$d$ row. We refer to this effect as the $d$-band filling rule.

In the present work we determine the electronic and magnetic properties of an Fe ML on the FCC(001), BCC(001) and HCP(0001) surfaces of 5$d$ transition metals using first principles calculations. In all considered cases we take into account the inward relaxation of the Fe ML. The layer relaxation was determined by using the Vienna Ab initio Simulation Package (VASP) \cite{kresseVASP1, kresseVASP2, hafnerVASP, kressePAW}. In order to get a prediction about the magnetic ground state of the Fe ML we calculate first principles spin model parameters. To obtain the model parameters we use the spin-cluster expansion technique combined with the relativistic disordered local moment (RDLM) scheme \cite{adeak,szunyoghSCE}. We study the influence of the isotropic Heisenberg couplings on the magnetic ground state as well as the role of Dzyaloshinskii-Moriya (DM) interactions representing a relativistic correction to the Heisenberg model.
The theoretical and computational details are briefly given in Section 2. In Sections 3 and 4 our results are presented on the magnetic properties and ground state magnetic structure of an Fe ML deposited on different 5$d$ transition metal surfaces. In particular, we find for the Ta substrate that the Fe out-of-plane relaxation itself causes a crossing from the FM to the AFM ground state. The frustration of the antiferromagnetic isotropic exchange interactions on HCP surfaces leads to a N\'eel state in the case of Fe/Re(0001), whereas, due to large DM interactions, an AFM spin spiral state is formed in Fe/Os(0001).

\section{Theory and computational details}

First principles descriptions of the magnetism of itinerant electron systems mostly rely on the adiabatic decoupling of fast electronic fluctuations characterized by typical hopping times of $~ 10^{-15}$~s from the
slow transversal motion of spins with time-scales related to inverse spin-wave frequencies of $~ 10^{-13}$~s \cite{gyorffy}.
As further simplification of the theory longitudinal spin-fluctuations are often neglected and
the so-called rigid spin approximation is used assigning each atom $i$ with a spin moment, $\vec{M}_i =M_i \vec{e}_{i}$ with $|\vec{e}_{i}|=1$ \cite{antropov}. A system of $N$ spin moments  can then be described by the set of orientations (spin configurations) $\{\vec{e}\}=\{\vec{e}_{1},\dots , \vec{e}_{N}\}$ and the ab initio grand potential of the itinerant electrons can be identified as an effective energy expression $E(\{\vec{e}\})$ of the classical spin system \cite{gyorffy}. The second order approximation to $E(\{\vec{e}\})$ refers to a generalized Heisenberg model,
\begin{equation}
E(\{\vec{e}\})=E_{0}+\sum_{i=1}^{N}\vec{e}_{i}\mathbf{K}_{i}\vec{e}_{i}-\frac{1}{2}\sum_{\substack{i,j=1 \\ (i\neq j)}}^{N}\vec{e}_{i}\mathbf{J}_{ij}\vec{e}_{j} \:, \label{GenHeis}
\end{equation}
where $\mathbf{K}_{i}$ and $\mathbf{J}_{ij}$ are the second-order anisotropy matrices and tensorial exchange interactions, respectively. The exchange coupling can be decomposed into an isotropic component, $J_{ij} \mathbf{I}$ with $J_{ij}=\frac{1}{3} \Tr\mathbf{J}_{ij}$, an antisymmetric component
$\mathbf{J}_{ij}^\mathrm{A}=\frac{1}{2} \left( \mathbf{J}_{ij}-\mathbf{J}_{ij}^\mathrm{T} \right)$,
and a traceless symmetric part
$\mathbf{J}_{ij}^\mathrm{S}=\frac{1}{2} \left( \mathbf{J}_{ij}+\mathbf{J}_{ij}^\mathrm{T} \right)-J_{ij} \mathbf{I}$,
where $\Tr$ and the superscript T label in order the trace and the transpose of a matrix,  and $\mathbf{I}$ is the unit matrix.  As is well-known, $J_{ij}$ is the scalar Heisenberg coupling and the energy term
$-\vec{e}_{i} \mathbf{J}_{ij}^\mathrm{A} \vec{e}_{j}=\vec{D}_{ij} \left( \vec{e}_{i} \times \vec{e}_{j} \right)$
corresponds to the Dzyaloshinskii-Moriya (DM) interaction with $\vec{D}_{ij}$ being the DM vector \cite{Dzyaloshinsky,Moriya}.  Furthermore, the magnetic anisotropy energy (MAE), defined as the energy difference  between two uniformly magnetized states of the system, $\vec{e}_i=\vec{e}_\alpha$ and  $\vec{e}_i=\vec{e}_\beta$
$(\forall i)$, can be expressed as a sum of on-site and two-site anisotropies, $\Delta E_\mathrm{os}$ and
 $\Delta E_\mathrm{ts}$,
\begin{equation}
\Delta E =E(\vec{e}_\alpha)-E(\vec{e}_\beta)  =
\Delta E_\mathrm{os} + \Delta E_\mathrm{ts} \: ,
\end{equation}
\begin{equation}
\Delta E_\mathrm{os} = \sum_{i=1}^{N} \left(
\vec{e}_\alpha \mathbf{K}_{i}\vec{e}_\alpha -  \vec{e}_\beta \mathbf{K}_{i}\vec{e}_\beta
\right)  \:, \label{dEos}
\end{equation}
\begin{equation}
\Delta E_\mathrm{ts} =-\frac{1}{2}\sum_{\substack{i,j=1 \\ (i\neq j)}}^{N} \left(
\vec{e}_\alpha\mathbf{J}^\mathrm{S}_{ij}\vec{e}_\alpha -
\vec{e}_\beta \mathbf{J}^\mathrm{S}_{ij}\vec{e}_\beta  \right) \:. \label{dEts}
\end{equation}

Neglecting self-consistent longitudinal spin-fluctuations, two methods are known in the literature for mapping the energy (grand-canonical potential) from first-principles calculations to the spin Hamiltonian in Eq.~(\ref{GenHeis}). The relativistic torque method \cite{udvardiRTM,ebertRTM} makes use of infinitesimal rotations around specific magnetic configurations, mostly, around  the ferromagnetic state. Deriving all elements of the matrices $\mathbf{K}_{i}$ and $\mathbf{J}_{ij}$ can, however, be very tedious in case of reduced symmetry within this method. Furthermore, for systems with ground states far from the FM state the derived interaction parameters might be inconsistent with the ground state. The spin-cluster expansion (SCE) technique
developed by Drautz and F\"ahnle \cite{DrautzSCE1, DrautzSCE2} provides a systematic parametrization of the adiabatic magnetic energy that, in principle, avoids the above problems. The crucial step of the SCE is the evaluation
of orientational averages of the grand potential restricted to fixed spin-configurations of selected clusters.
These restricted averages can efficiently be elucidated using mean field approximation.
The disordered local moment (DLM) scheme of the density functional theory \cite{gyorffy} is an
extremely useful mean field  description of a magnetic system. The DLM was implemented within the Korringa-Kohn-Rostoker (KKR) electronic structure method by Gy\"{o}rffy et al.\ \cite{gyorffy} and extended to relativistic electron theory by Staunton et al.\ \cite{stauntonRDLM1, stauntonRDLM2, buruzsRDLM}. Lately, the SCE method was combined with the relativistic disordered local moment (RDLM) scheme to determine the parameters of the spin Hamiltonian in Eq.~(\ref{GenHeis}) \cite{adeak,szunyoghSCE}. It should be noted that within the SCE-RDLM scheme the spin-spin correlations in the paramagnetic state of the system are calculated.

The structural relaxation of the Fe monolayer was considered for all studied (FCC, BCC, HCP) substrates.
We performed geometry relaxation based on the density functional theory (DFT) within the generalized gradient approximation (GGA)
implemented in the Vienna Ab-initio Simulation Package (VASP) \cite{kresseVASP1, kresseVASP2, hafnerVASP}.
We employed a plane-wave basis set for the electronic wave function expansion, and the projector-augmented wave (PAW) method
\cite{kressePAW} for the description of the electron-ion interaction. Moreover, the Perdew-Wang (PW91) parametrization of the
exchange-correlation functional \cite{perdew} was used.
We modelled the Fe/5$d$ systems by a slab of seven layers (1 ML Fe + 6 ML 5$d$), where the two topmost layers were freely relaxed
in the out-of-plane direction. We used the experimental lattice constants of the substrates and an $11\times 11\times 1$
Monkhorst-Pack k-point grid for sampling the Brillouin zone (BZ).

Taking the relaxed geometries derived by using VASP we performed self-consistent calculations for the Fe monolayer placed on the semi-infinite 5$d$ substrates in terms of the screened Korringa-Kohn-Rostoker (SKKR) method \cite{szunyoghKKR, zellerKKR}. We employed the scalar relativistic DLM approach \cite{gyorffy} resulting in a paramagnetic configuration because we did not have \textit{a priori} information about the magnetic ground state of the Fe monolayers. The local spin-density approximation (LSDA) as parametrized by Vosko et al.\ \cite{vosko} was applied. The effective potentials and fields were treated within the atomic sphere approximation (ASA) with an angular momentum cut-off of $l_{max}=3$. The energy integrals were performed by sampling $16$ points on a semicircle contour in the upper complex semi-plane. Following the self-consistent calculations, we used the SCE-RDLM method as described above to calculate the parameters of the spin Hamiltonian.


\section{Geometrical structure and magnetic moments}

For all considered Fe monolayer systems, Table \ref{moment-data} summarizes the main structural data, namely, the in-plane lattice constant of the substrate $a$, the interlayer distance between the Fe monolayer and the top substrate layer $d$, the relative Fe layer relaxation compared to the bulk substrate interlayer distance $\delta$ obtained by using VASP, together with the Fe spin magnetic moments $m_{Fe}$, and the magnetic anisotropy energies.
For the latter ones, we employed the SCE-RDLM method and considered the normal-to-plane $z$ and in-plane $x$ directions in getting
the on-site magnetic anisotropy energy $\Delta E_\mathrm{os}$ and the two-site magnetic anisotropy energy $\Delta E_\mathrm{ts}$ in Eqs.~(\ref{dEos}) and (\ref{dEts}), respectively. Also shown is the total magnetic anisotropy energy of the system, $\Delta E = \Delta E_\mathrm{os} + \Delta E_\mathrm{ts}$.
In our definition the negative sign of the magnetic anisotropy energy means that the easy axis is perpendicular to the surface plane.

\begin{table}[htb!]
\begin{center}
\begin{tabular}{c c c c c c c c}
\hline
            & a   &   d   & $\delta$ &  $m_{Fe}$   & $\Delta E_\mathrm{os}$ &  $\Delta E_\mathrm{ts}$ &$\Delta E$  \\
            &(\AA)& (\AA) &    (\%)   & ($\mu_{B}$) &  (meV)  &     (meV)  & (meV)   \\
\hline
\hline
Fe/Ir(001)  & 2.71  & 1.74 & -9.1 & 2.94 &  0.15 &  0.18 &  0.33 \\
Fe/Pt(001)  & 2.77  & 1.73 &-11.7 & 3.26 & -0.06 &  1.10 &  1.04 \\
Fe/Au(001)  & 2.88  & 1.76 &-13.6 & 3.33 & -0.50 & -0.43 & -0.93 \\
\hline
Fe/Ta(001)  & 3.30  & 1.09 &-33.8 & 1.58 & -0.52 &  0.22 & -0.30 \\
Fe/W(001)   & 3.16  & 1.25 &-21.4 & 1.72 &  0.04 & -1.71 & -1.67 \\
\hline
Fe/Hf(0001) & 3.19  & 1.68 &-24.9 & 2.71 & -1.92 & -0.07 & -1.99 \\
Fe/Re(0001) & 2.76  & 1.89 &-14.6 & 2.37 &  0.73 &  2.03 &  2.76 \\
Fe/Os(0001) & 2.73  & 2.03 & -5.6 & 2.60 &  0.52 &  0.97 &  1.49 \\
\hline
\hline
\end{tabular}
\caption{Computational results on structural data of an Fe monolayer on 5$d$ substrates ($a$ denotes the in-plane lattice constant of the substrate, $d$ is the interlayer distance between the topmost substrate atomic layer and the Fe overlayer, $\delta$ is the ratio of $d$ and the bulk interlayer distance minus one, thus negative $\delta$ means inward relaxation), the spin magnetic moment of Fe, $m_{Fe}$ obtained from the DLM calculations, the on-site, the two-site and the total magnetic anisotropy energies, $\Delta E_\mathrm{os}$, $\Delta E_\mathrm{ts}$, and $\Delta E$, respectively.}
\label{moment-data}
\end{center}
\end{table}

In Table~\ref{moment-data} we separate three surface types: in case of Ir, Pt, and Au we considered the (001) surface of the FCC lattice, in case of Ta and W the BCC(001) surface, and for Hf, Re, and Os the HCP(0001) surface. For the FCC(001) substrates we find that the $d$ interlayer distance is fairly constant, and the absolute value of $\delta$ increases with increasing $a$. Concomitantly, the spin magnetic moment of Fe monotonically increases from Ir towards Au due to the increasing $d$-band filling of the substrate \cite{hardrat}. Because of the open atomic structure of the BCC(001) surface, the Fe layer relaxation is relatively large for BCC(001) substrates, particularly for Ta.
Since Ta and W are the first two 5$d$ elements in the periodic table after Hf, a smaller Fe magnetic moment is expected for these substrates than for Hf according to the $d$-band filling rule. This effect is even amplified by the increased hybridization between the Fe layer and the substrate layer due to the large layer relaxation mentioned above, therefore, we find the smallest Fe magnetic moments in case of the BCC substrates.

\begin{figure}[t!]
\begin{center}
\includegraphics[scale=1.7]{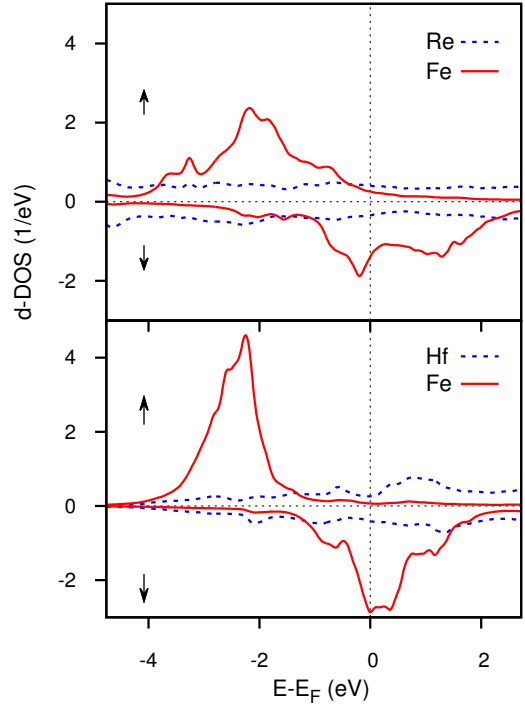}
\caption{Top: Calculated spin-polarized partial $d$-like densities of states ($d$-DOS) projected on the Fe atom and on the Re atom in the topmost Re layer in the Fe/Re(0001) system. Bottom: The same as in the top figure for the Fe/Hf(0001).}
\label{dos}
\end{center}
\end{figure}

In case of the HCP substrates we observe an interesting trend for the Fe magnetic moment. 
As can be seen from Table~\ref{moment-data} for the Hf substrate, the Fe layer relaxation is quite large ($-25\%$) and the Fe magnetic moment is also relatively large, 2.71~$\mu_B$. This can clearly be related to the very low $d$-band filling of Hf.
In case of the Re substrate the relative inward relaxation of the Fe layer $\delta$ is significantly smaller than in case of Hf, however, due to the larger
$d$-band filling of Re, the Fe magnetic moment is decreased. 
In Fig.~\ref{dos} we demonstrate the above effects by comparing the calculated spin-polarized partial $d$-like densities of states projected on the Fe and the topmost substrate layers for the cases of Hf and Re substrates. From Fig.~\ref{dos} it is obvious that Hf has less $d$-states below the Fermi level than Re, and thus the hybridization between the Fe and the topmost Hf layer is relatively small despite the large inward Fe layer relaxation. Re has more $d$-states below the Fermi level and the hybridization between Fe and Re is remarkably larger.
Comparing the cases of Re and Os substrates we note that the Fe-Re interlayer distance is smaller than the Fe-Os interlayer distance, therefore, a reduced hybridization is expected for the Fe/Os(0001) system, explaining the increased Fe moment on the Os substrate.

\section{Spin-interactions and magnetic ground state of the Fe monolayers}

\subsection{Fe on FCC substrates}

\begin{figure}[ht!]
\begin{center}
\includegraphics[scale=0.7]{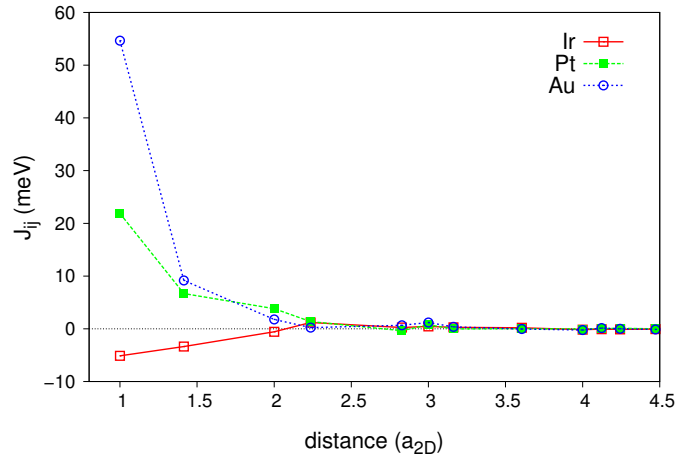}
\caption{Calculated Fe-Fe isotropic exchange interactions as a function of the inter-atomic distance in units of the in-plane lattice constant for the Fe monolayer on FCC(001) substrates.}
\label{fcc-jij}
\end{center}
\end{figure}

In Fig.~\ref{fcc-jij} the Fe-Fe isotropic exchange interactions are shown as a function of the inter-atomic distance for the Fe monolayer on FCC(001) substrates. Using the Hamiltonian in Eq.~(\ref{GenHeis}), a positive sign of the exchange parameter means ferromagnetic coupling, while negative sign refers to the antiferromagnetic coupling. For the Au and Pt substrates all couplings are ferromagnetic (FM) leading to a stable FM state for the Fe monolayer on these substrates. The calculated magnetic anisotropy energies, see Table~1, imply a normal-to-plane direction of the magnetization for Au substrate and an in-plane orientation for Pt substrate. Our result for Fe/Au(001) is in line with that of Ref.~\cite{feau}, though the magnetic anisotropy energy is likely magnified by the layer relaxation included in the present study. In contradiction to our present result, in Ref.~\cite{fept} a weak out-of-plane magnetic anisotropy was reported for Fe/Pt(001).
This difference in the MAE can be attributed to the fact that our results are obtained from the exchange interactions calculated in the DLM state taking a semi-infinite Pt geometry, whereas in Ref.~\cite{fept} a total energy difference was calculated in a thin slab geometry.
  
In case of Ir substrate, we find antiferromagnetic (AFM) Fe couplings for the first three shells. This finding is in good agreement with Ref.\ \cite{adeak}. It is also apparent that the magnitude of the exchange interaction in the first shell is much smaller for Ir than for Au and Pt substrates. Additionally, the second and third nearest neighbour AFM couplings are comparable in size with the nearest neighbour AFM coupling. These frustrated AFM couplings give rise to complex magnetic ground states for the Fe/Ir(001) overlayer \cite{adeak, kudrnovsky}.

\subsection{Fe on BCC substrates}

Ta(001) and W(001) surfaces have been the focus of a number of theoretical and experimental studies \cite{kubetzka, qian, ferriani, ondracek}. In case of an Fe monolayer on W(001) surface the magnetic ground state is antiferromagnetic, while for the Fe ML on Ta(001) substrate the ground state is ferromagnetic. By manufacturing alloys in different stoichiometries from these two substrate elements, a crossing can be found from the AFM to the FM state of the Fe monolayer \cite{ondracek}. It should be noted that in previous studies \cite{kubetzka, ondracek} the Fe layer relaxation on the Ta substrate was assumed to be the same as on W substrate with the reasoning that Ta and W are neighbouring elements in the periodic table. In contrary, we found a considerable difference in the out-of-plane relaxation of the Fe ML on these two substrates, see Table \ref{moment-data}. 

\begin{figure}[ht!]
\begin{center}
\includegraphics[scale=0.7]{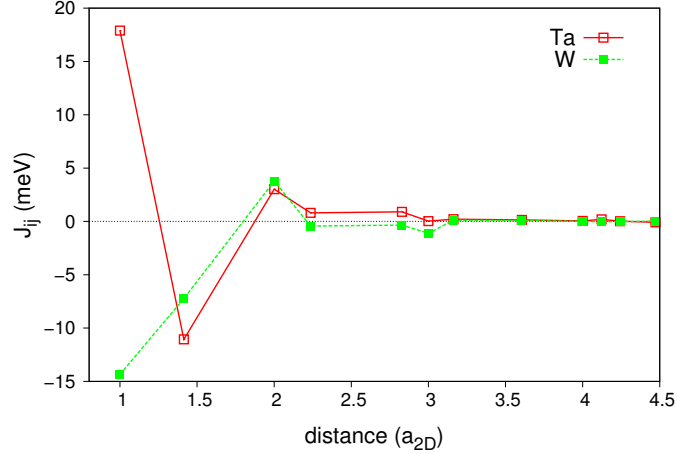}
\caption{Calculated Fe-Fe isotropic exchange interactions as a function of the inter-atomic distance in units of the in-plane lattice constant for the Fe monolayer on BCC(001) substrates.}
\label{bcc-jij}
\end{center}
\end{figure}

The calculated isotropic Fe-Fe exchange interactions are shown in Fig.~\ref{bcc-jij}. As can be seen, on the W substrate the nearest and next nearest neighbour interactions are antiferromagnetic, in the third shell the exchange interaction is FM that turns back to AFM in the fourth and fifth shells. For the Ta substrate the nearest neighbour coupling is FM, whereas the next nearest neighbour interaction becomes AFM, comparable in magnitude with the nearest neighbour one. From shells three to five the interactions are ferromagnetic. All these findings are similar to those reported in Ref.\ \cite{ondracek}. However, mainly due to the sizeable AFM coupling in the second shell ($J_2$= -11.06 meV), the magnetic ground state is not obvious for the Fe/Ta(001) system.

To obtain an estimate for the magnetic ground state, we calculated the Fourier transform of the tensorial coupling matrices, $\mathbf{J}(\vec{q})$.
For a spin-system that is described by a Heisenberg Hamiltonian, Eq.~(\ref{GenHeis}), the energy of a spin spiral state with momentum vector
$\vec{q}$ is given by the minimum eigenvalue of -$\mathbf{J}(\vec{q})$, or equivalently, by the maximum eigenvalue of $\mathbf{J}(\vec{q})$  \cite{ondracek,adeak,kudrnovsky}, which we shall denote by $J(\vec{q})$ in the following. Taking the maximum value of $J(\vec{q})$ over the $\vec{q}$ vectors in the Brillouin zone, a suitable approximation for the
magnetic ground state can be obtained.   
A maximum at the center of the Brillouin zone, i.e. at the $\overline{\Gamma}$ point means a ferromagnetic ground state, whilst a maximum located at a general $\vec{q}$ vector of the Brillouin zone corresponds to a more complex magnetic ground state including AFM or spin spiral states. For instance if the maximum of $J(\vec{q})$ is at the $\overline{\mathrm{M}}=(1,1)\frac{\pi}{a_{2D}}$ point of the Brillouin zone of a 2D square lattice ($a_{2D}$ being the 2D lattice constant) then it indicates a $c(2\times2)$ (checkerboard) AFM ground state. 
Similarly, a maximum at the $\overline{\mathrm{X}}=(1,0)\frac{\pi}{a_{2D}}$ point corresponds to the $p(2\times1)$ (row-wise) AFM ground state.

\begin{figure}[ht!]
\begin{center}
\includegraphics[scale=0.7]{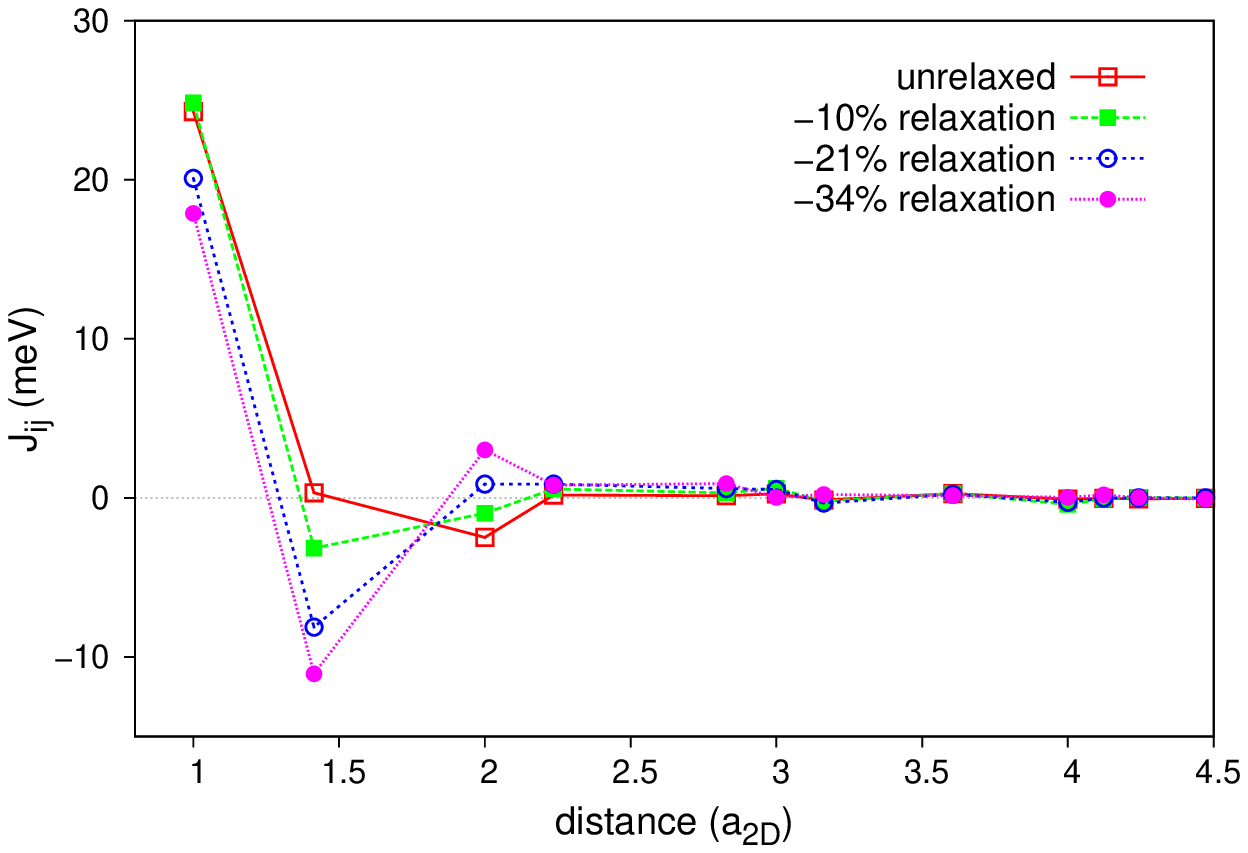}
\includegraphics[scale=0.7]{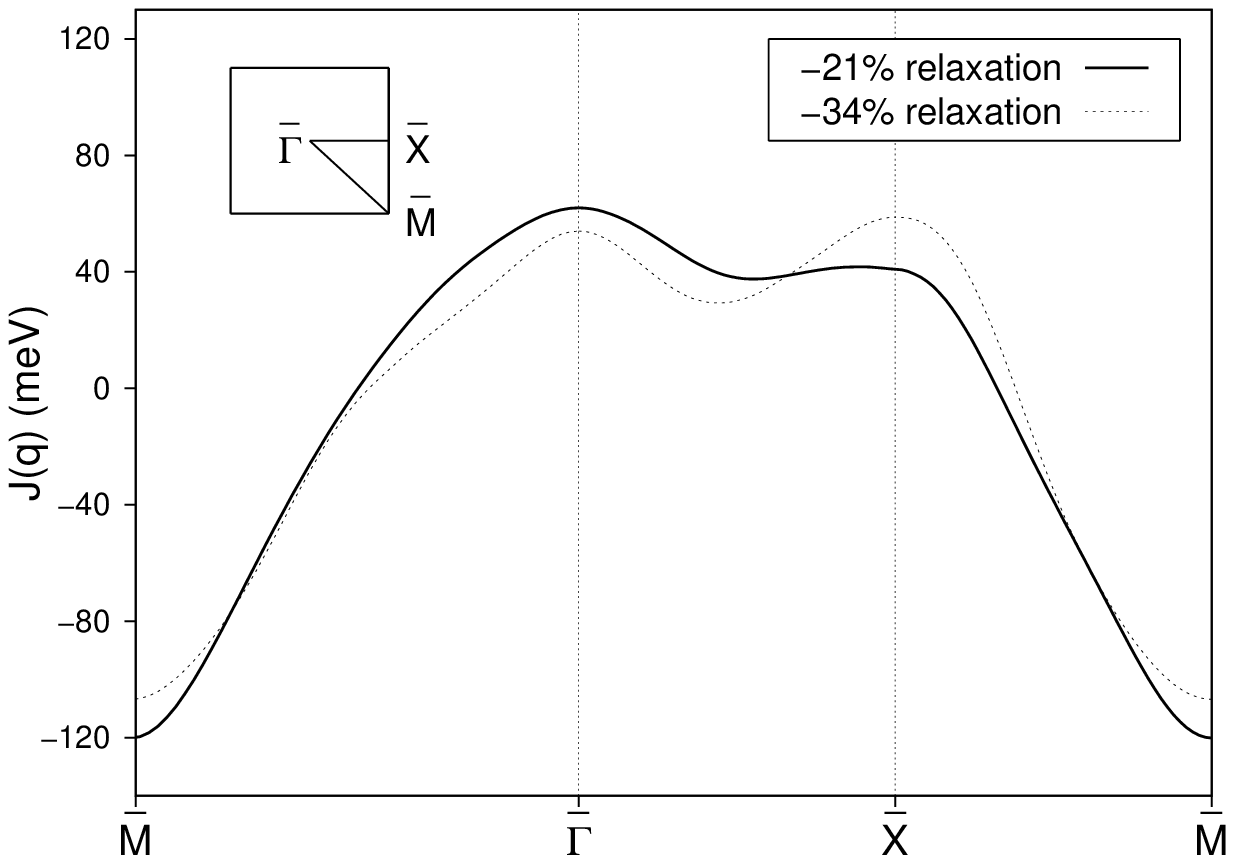}
\caption{Top: Calculated Fe-Fe isotropic exchange interactions for various relaxations of an Fe layer on Ta(001) surface as a function of the inter-atomic distance. Bottom: maximum eigenvalues of the Fourier transform of the tensorial coupling matrix, $J(\vec{q})$, along high-symmetry lines of the Brillouin zone considering two different Fe layer relaxations.}
\label{relax-ta}
\end{center}
\end{figure}

Based on our calculations of $J(\vec{q})$, we find that for the Fe/Ta(001) system in the relaxed structure the ground state is row-wise AFM. Taking, however, a smaller inward Fe layer relaxation the magnetic ground state of the Fe overlayer on Ta(001) becomes increasingly FM. Fig. \ref{relax-ta} shows that with increasing inward Fe layer relaxation the nearest neighbour FM coupling decreases and the next nearest neighbour AFM coupling is enhanced. It is also worthwhile to note that the third nearest neighbour coupling turns from AFM to FM with increasing inward Fe layer relaxation. Moreover, as can be seen in Fig. \ref{relax-ta}, taking the same relative Fe layer relaxation as was obtained in the Fe/W(001) system (-21\%), $J(\vec{q})$ reaches the maximum at the $\overline{\Gamma}$ point which implies a FM ground state. On the other hand, the maximum of $J(\vec{q})$ is found at the $\overline{\mathrm{X}}$ point by considering the calculated relative relaxation (-34\%) for the Fe/Ta(001) system that corresponds to a row-wise AFM ground state. 
This result shows the importance of the relaxations of the geometry for determining the magnetic ground state. Performing total energy calculations for the relaxed Fe/Ta(001) system by VASP we found that the row-wise AFM state has the smallest total energy compared to the other two magnetic structures, namely, the $c(2\times2)$ AFM and the FM states with 77 meV/Fe atom and 9 meV/Fe atom larger total energies, respectively. For the relaxed Fe/W(001) system only the third nearest neighbour Fe coupling is FM, and all other couplings are AFM, as can be seen in Fig.\ \ref{bcc-jij}. The maximum of $J(\vec{q})$ is obtained at the $\overline{\mathrm{M}}$ point of the BZ which corresponds to the $c(2\times2)$ AFM magnetic state. This ordering is very stable against larger inward Fe layer relaxations (not shown), similar to the findings in Ref.\ \cite{kubetzka}.

\subsection{Fe on HCP substrates}

\begin{figure}[t!]
\begin{center}
\includegraphics[scale=0.7]{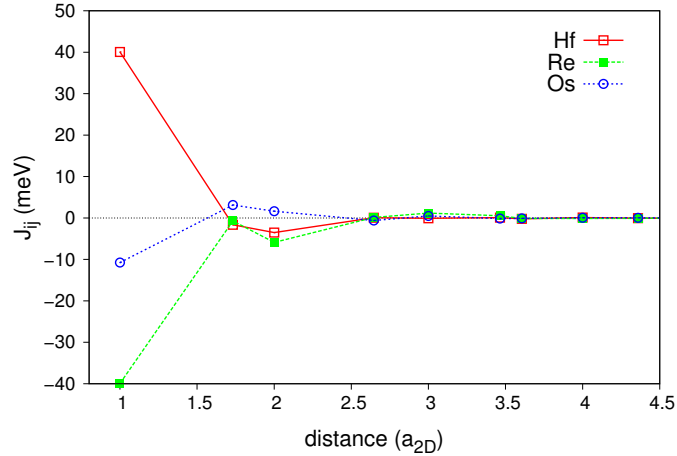}
\caption{Calculated Fe-Fe isotropic exchange interactions as a function of the inter-atomic distance in units of the in-plane lattice constant for the Fe monolayer on HCP(0001) substrates.}
\label{hcp-jij}
\end{center}
\end{figure}

\begin{figure}[ht!]
\begin{center}
\includegraphics[scale=0.7]{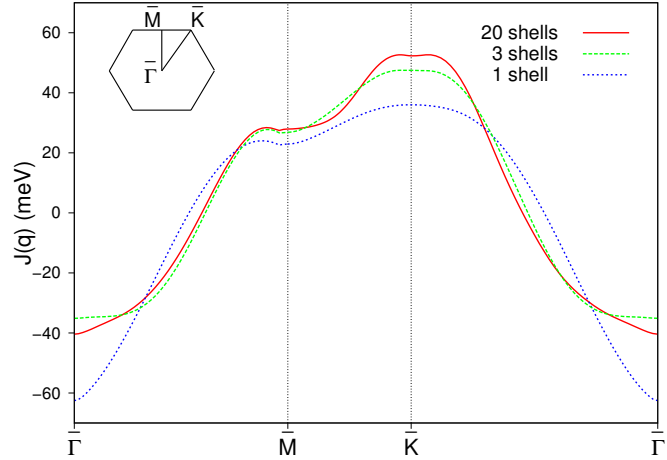}
\caption{Maximum eigenvalues of the Fourier transform of the tensorial exchange matrix, $J(\vec{q})$, along high-symmetry lines of the Brillouin zone, calculated by including different number of shells for the Fe/Os(0001) system.}
\label{os-jq}
\end{center}
\end{figure}

More complex behaviour of the isotropic Fe-Fe interactions are found for  HCP(0001) substrates. As can be seen in Fig.\ \ref{hcp-jij}, for the Hf substrate the nearest neighbour Fe coupling is strongly ferromagnetic, and all couplings at larger inter-atomic distances of AFM character are negligible as compared to the nearest neighbour exchange interaction. Therefore, the magnetic ground state is likely ferromagnetic. A different situation can be observed for the other two considered substrates (Re, Os), where the nearest neighbour couplings are AFM, whilst the second and third nearest neighbour couplings are AFM for Fe/Re(0001) and FM for Fe/Os(0001). The dominant nearest neighbour AFM interaction on a hexagonal lattice leads to the frustration of magnetic moments, thus, it is the origin for complex magnetic states. For the Fe/Re(0001) system the maximum of $J(\vec{q})$ is found at the $\overline{\mathrm{K}}$ point of the BZ which corresponds to a $120^{\circ}$ N\'eel state. In case of Fe/Os(0001) the maximum of $J(\vec{q})$ is around $\vec{q}=(0.62,1.08)\frac{\pi}{a_{2D}}$
that is, very close to the boundary of the two dimensional BZ, implying that the magnetic ground state can be a spin spiral modulation of the $120^{\circ}$ N\'eel state. The result of $J(\vec{q})$ for the Fe/Os(0001) system is reported in Fig.\ \ref{os-jq} considering exchange interactions for different number of shells in the determination of $J(\vec{q})$. We find that by considering one shell only, the maximum of $J(\vec{q})$ is exactly found at the $\overline{\mathrm{K}}$ point of the BZ, whereas by taking an increased number of shells the maximum is shifted away from the $\overline{\mathrm{K}}$ point, thus the magnetic ground state is sensitive to the number of shells included in the evaluation of $J(\vec{q})$.

In particular cases, where the magnitudes of the isotropic couplings are small, it is possible that Dzyaloshinskii-Moriya (DM) interactions play an important role in the formation 
of the magnetic ground state \cite{Bode-MnW, Ferriani-MnW, Udvardi-MnW}. A good qualitative measure to classify spin spiral like magnetic patterns is
the ratio of the magnitudes of the nearest neighbour DM vector ($D$) and isotropic exchange interaction ($J$). In Table \ref{DM-data} this ratio is shown for all considered Fe monolayers together with the value of $J$ and the character of the obtained magnetic ground state. It can be seen that, among the FCC(001) substrates, the $D/J$ ratio for Au is practically negligible, whereas for Ir and Pt $D/J \simeq 0.2$. The Fe monolayer on W, Ta, and Hf substrates also exhibits very small $D/J$ ratios, for Fe on Re it is again sizeable and in case of the Os(0001) host it has the largest value, $D/J \simeq 0.455$.
\begin{table}[ht!]
\begin{center}
\begin{tabular}{c|c|c|c}
\hline
Substrates & $D/J$ & $J$ (meV) & Ground state\\
\hline
Ir(001)    &    0.196  & -5.13 & spin spiral \\
Pt(001)    &    0.200  & 21.88 & FM \\
Au(001)    &    0.0003 & 54.65 & FM \\
Ta(001)    &    0.039  & 17.89 & AFM \\
 W(001)    &    0.031  &-14.34 & AFM \\
Hf(0001)   &    0.054  & 40.05 & FM \\
Re(0001)   &    0.149  &-39.92 & N\'eel-AFM \\
Os(0001)   &    0.455  &-10.76 & AFM spin spiral \\
\hline
\hline
\end{tabular}
\caption{Ratio of the magnitudes of the Fe-Fe nearest neighbour DM vector ($D$) and the nearest neighbour isotropic exchange coupling ($J$) for all considered Fe overlayer systems. The values of $J$ and the character of the obtained magnetic ground states are also shown. Note that for Fe/Ta(001) the second nearest neighbour Fe-Fe isotropic exchange ($J_2$= -11.06 meV) is responsible for the AFM ground state.}
\label{DM-data}
\end{center}
\end{table}

For finding more complex magnetic ground states, we performed zero temperature Landau-Lifshitz-Gilbert spin dynamics simulations. We used a two-dimensional lattice of $128\times128$ sites with free boundary conditions and the full tensorial exchange interactions were considered, i.e., including the isotropic, the DM and the two-site anisotropy terms. Each simulation was initialized at a random spin configuration and continued until the absolute difference in the energy of the spin-system between two steps reached the value of $10^{-5}$ mRy. Here, we report results on the Fe/Re(0001) and Fe/Os(0001) systems only.

\begin{figure}[ht!]
\begin{center}
\includegraphics[scale=0.3]{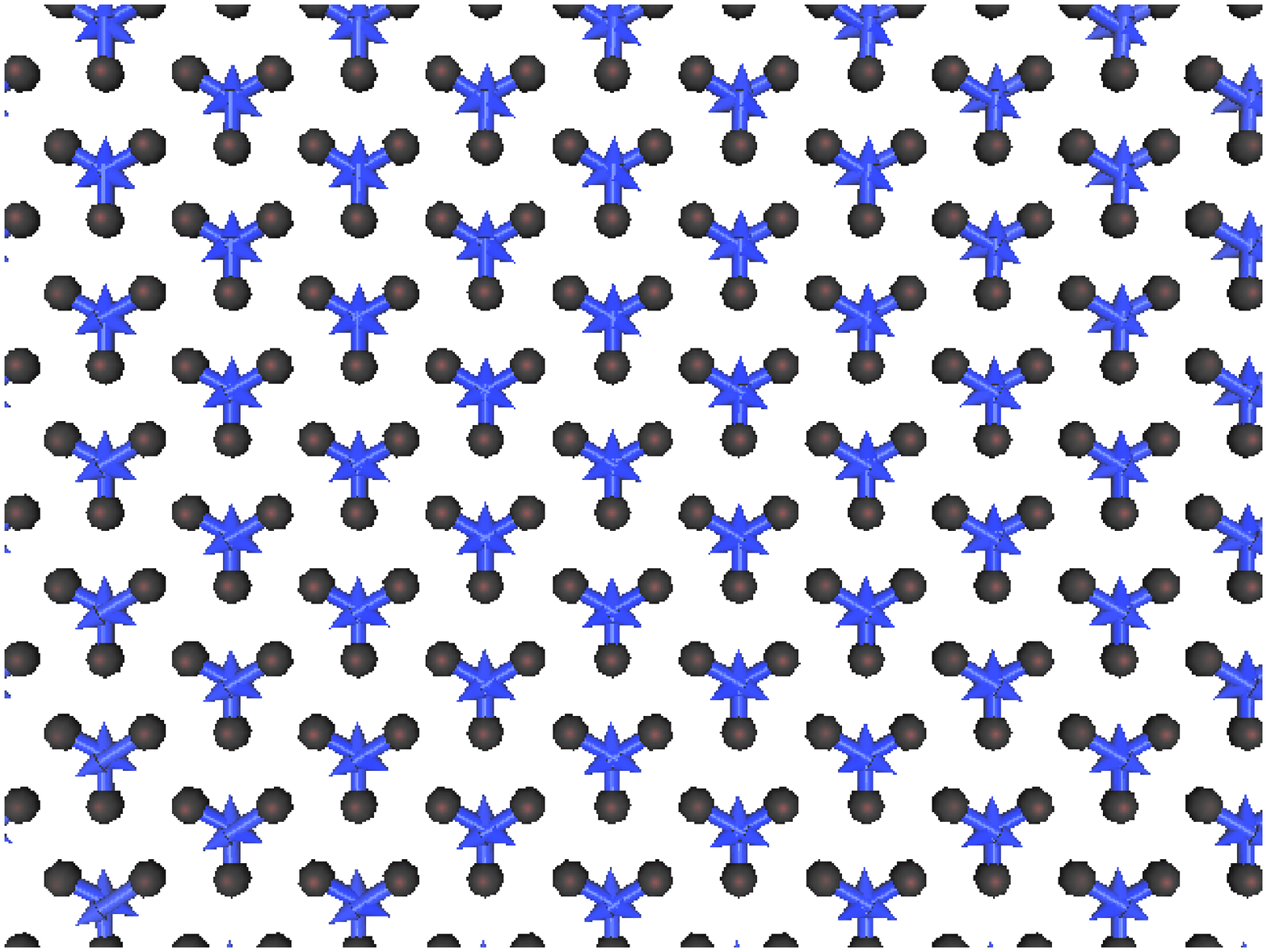} \hskip 0.5cm
\includegraphics[scale=0.3]{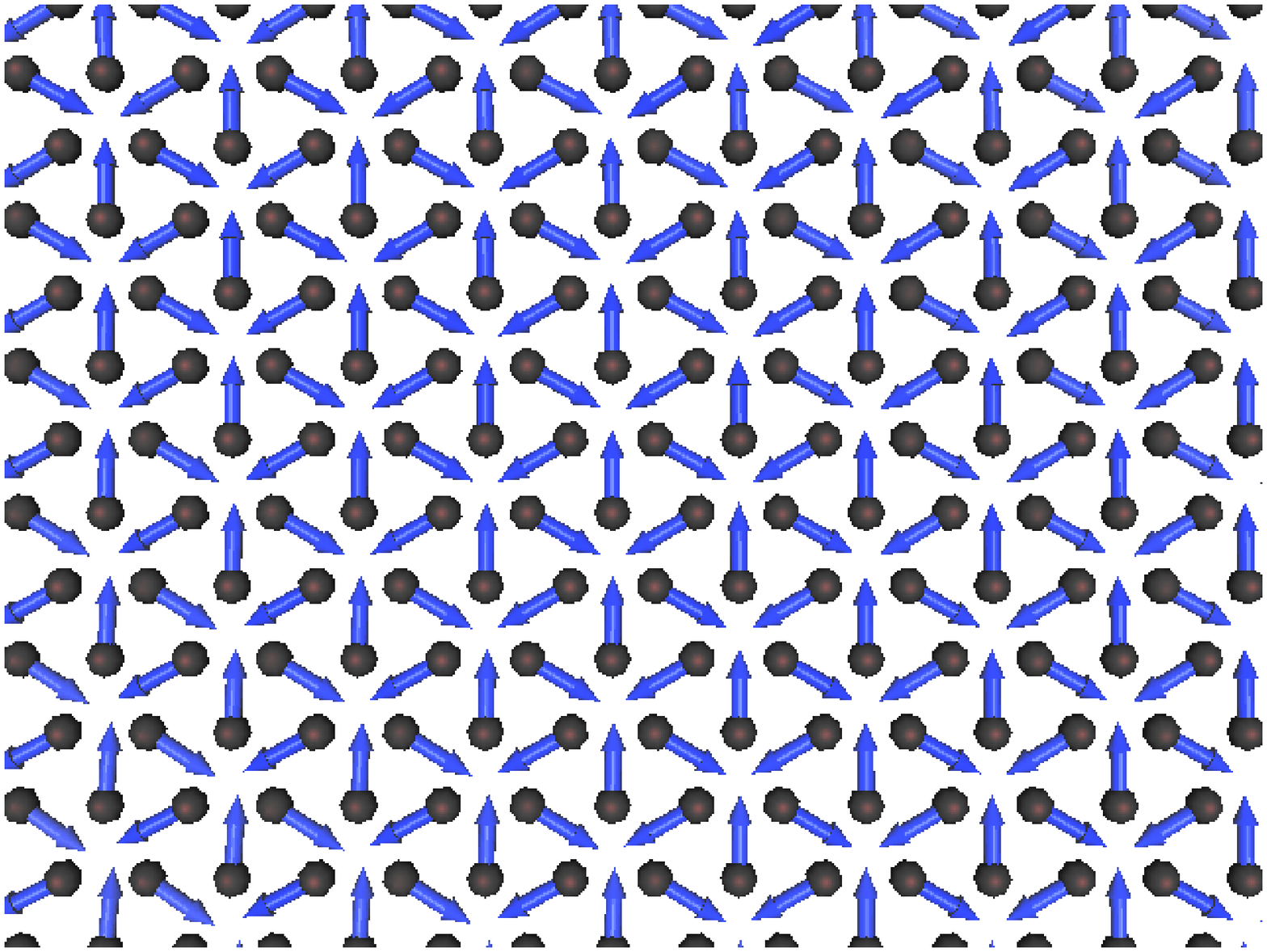}
\caption{$120^{\circ}$ N\'eel type AFM ground state spin-configuration of the Fe monolayer on the Re(0001) substrate in case of positive (left side) and negative (right side) magnetic chirality.}
\label{sd-re}
\end{center}
\end{figure}

From the spin dynamics simulations we obtained a 120$^\circ$ N\'eel-type AFM state as the ground state for the Fe/Re(0001) system, which agrees with a recent experimental finding \cite{ouazi14}. This result is in accordance with the estimation based on $J(\vec{q})$ that yielded the $\overline{\mathrm{K}}$ point considering tensorial exchange interaction.
The $120^{\circ}$ N\'eel states can occur with two possible magnetic chiralities, in a manner similar to the case of Cr trimers \cite{antala}, and a Cr monolayer \cite{palotasCr}.
As can be inferred from Fig.\ \ref{sd-re}, the magnetic configuration of one chirality can be obtained from the other if the magnetic directions of two atoms in the triangular magnetic unit cell are interchanged, while the magnetic direction of the third atom remains unchanged. We note that positive/negative chirality of the 120$^\circ$ N\'eel-type AFM state corresponds to the total energy minimum of the system at the $+/- \overline{\mathrm{K}}$ point of the Brillouin zone.
It is well-known that in the case of isotropic Heisenberg spin-model, even including anisotropy terms, the two chiral states are degenerate
and it is the Dzyaloshinskii-Moriya interaction that lifts this degeneracy \cite{antala}.
From the spin-dynamics simulations we find that the magnetic ground state is of positive chirality and the energy difference between the positive and negative chirality states is $\Delta E=3.4$ meV/atom. 

\begin{figure}[ht!]
\begin{center}
\includegraphics[scale=0.35]{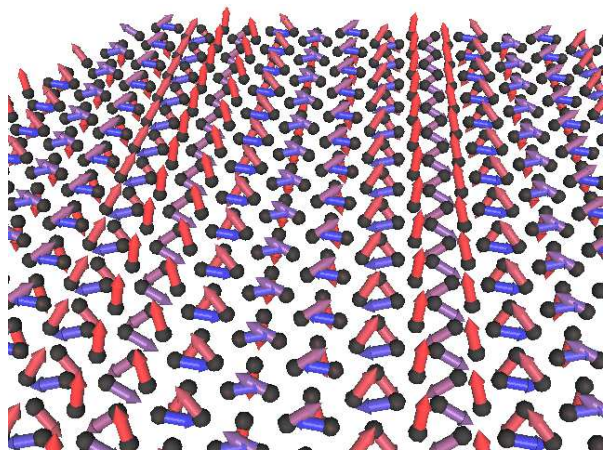}
\caption{Ground state AFM spin spiral configuration of the Fe monolayer on the Os(0001) substrate as obtained from spin-dynamics simulations.}
\label{sd-os}
\end{center}
\end{figure}

For the Fe/Os(0001) system, AFM nearest neighbour isotropic coupling and FM second and third nearest neighbour couplings can be found in Fig.\ \ref{hcp-jij}. The estimation of the ground state from $J(\vec{q})$ using tensorial exchange interactions resulted in a spin spiral state close to the 120$^\circ$ N\'eel-state: the maximum of $J(\vec{q})$ is near the $\overline{\mathrm{K}}$ point, at $\vec{q}=(0.62,1.08)\frac{\pi}{a_{2D}}$, see Fig.\ \ref{os-jq}.
The spin-configuration obtained from the spin dynamics simulations is depicted in Fig.\ \ref{sd-os}. Performing the Fourier transform of the spin-vectors on the lattice, we observed a well-defined peak at $\vec{q}=(0.63,1.08)\frac{\pi}{a_{2D}}$. Thus, the agreement for the magnetic ground state between the estimation based on $J(\vec{q})$ and on the spin dynamics simulations is remarkably good.
This $\vec{q}$ corresponds to a spin spiral wave length of $\lambda=1.6\cdot a_{2D}$, with the angle between the spin spiral propagation and the $x$ (nearest neighbour) direction being approximately $60^{\circ}$. It should be noted that due to the $C_{3v}$ symmetry of the lattice, there are three different propagation directions representing three degenerate spin spirals states. This leads to magnetic domain formations in the Fe monolayer that can be seen in spin-dynamics simulations on a larger length-scale.

\section{Conclusions}

We investigated magnetic properties and ordering of an Fe monolayer on 5$d$ substrates using first principles methods. We used the VASP code to obtain the inward Fe layer relaxations on the considered substrates, and calculated the atomic magnetic moments of the Fe monolayer by using the SKKR method. We employed the SCE-RDLM method to obtain exchange interaction parameters. Using these parameters we investigated the magnetic ground states of the Fe monolayer within an approximation based on the Fourier transform of the exchange interactions and, in selected cases, employing spin dynamics simulations.

The formation of the Fe magnetic moments was correlated with the geometrical relaxation of the Fe monolayer and the $d$-band filling of the substrate. Inspecting the isotropic exchange interactions, ferromagnetic ground states were concluded for FCC Au(001) and Pt(001) substrates, and a complex frustrated spin-structure for Ir(001) substrate. We found that for the BCC Ta(001) substrate the inward Fe layer relaxation is relatively large and thus the magnetic ground state becomes antiferromagnetic, whereas assuming a layer relaxation as obtained for Fe/W(001) the ground state is ferromagnetic. For the case of HCP Os(0001) substrate, we demonstrated that the range of the exchange interactions that are considered plays a crucial role in the determination of the magnetic ground state. The large antiferromagnetic nearest neighbour isotropic exchange interactions lead to a frustrated $120^{\circ}$ N\'eel state for the Fe monolayer on the Re(0001) substrate. By taking into account
the Dzyaloshinskii-Moriya interactions we determined the energetically favored magnetic chirality of the N\'eel state. The largest effect of the DM interactions was found for the Fe/Os(0001) system, where a spin spiral modulation of the N\'eel state occurred.

\section*{Acknowledgments}

This work was supported by the Hungarian Scientific Research Fund projects K77771, PD83353, K84078, and K91219. KP acknowledges the Bolyai Research Grant of the Hungarian Academy of Sciences. The work of LS was supported by the European Union, co-financed by the
European Social Fund, in the framework of T\'AMOP 4.2.4.A/2-11-1-2012-0001 National Excellence Program.
Support for the work of GMS was provided by the U.S. Department of Energy, Office of Energy Efficiency and
Renewable Energy (EERE), under its Vehicle Technologies Program, through the Ames
Laboratory. Ames Laboratory is operated by Iowa State University under contract DE-AC02-07CH11358. Partial support of the work of BU was from the same source.

\end{document}